\documentclass[openacc]{rstransa}

\usepackage{amsmath}
\usepackage{amssymb}
\usepackage{graphicx}

\usepackage{siunitx}
\usepackage[numbers,sort&compress]{natbib}

\newcommand*{\doi}[1]{\href{http://dx.doi.org/#1}{doi: #1}}

\usepackage{hyperref}
\usepackage{enumitem}

\usepackage{tcolorbox}
\definecolor{abstract-color}{cmyk}{0.04, 0.04, 0.12, 0.08}

\begin{document}

\title{An overall view of temperature oscillations in the solar chromosphere with ALMA}

\author{
S.~Jafarzadeh$^{1,2}$, S.~Wedemeyer$^{1,2}$, B.~Fleck$^{3}$, M.~Stangalini$^{4}$, D.~B.~Jess$^{5,6}$, R.~J.~Morton$^{7}$, M.~Szydlarski$^{1,2}$, V.~M.~J.~Henriques$^{1,2}$, X.~Zhu$^{8}$, T.~Wiegelmann$^{8}$, J.~C.~Guevara~G\'omez$^{1,2}$, S.~D.~T.~Grant$^{5}$, B.~Chen$^{9}$, K.~Reardon$^{10}$, and S.~M. White$^{11}$
}

\address{$^{1}$Rosseland Centre for Solar Physics, University of Oslo, P.O. Box 1029 Blindern, NO-0315 Oslo, Norway\\
$^{2}$Institute of Theoretical Astrophysics, University of Oslo, P.O. Box 1029 Blindern, NO-0315 Oslo, Norway\\
$^{3}$ESA Science and Operations Department, c/o NASA Goddard Space Flight Center, Greenbelt, MD 20771, USA\\
$^{4}$ASI Italian Space Agency, Via del Politecnico snc, I-00133 Rome, Italy\\
$^{5}$Astrophysics Research Centre, School of Mathematics and Physics, Queen’s University Belfast, Belfast, BT7 1NN, UK\\
$^{6}$Department of Physics and Astronomy, California State University Northridge, Northridge, CA 91330, USA\\
$^{7}$Department of Mathematics, Physics and Electrical Engineering, Northumbria University, Newcastle upon Tyne, NE1 8ST, UK\\
$^{8}$Max Planck Institute for Solar System Research, Justus-von-Liebig-Weg 3, D-37077 G{\"o}ttingen, Germany\\
$^{9}$Center for Solar-Terrestrial Research, New Jersey Institute of Technology, 323 M L King Jr. Blvd., Newark, NJ 07102-1982, USA\\
$^{10}$National Solar Observatory, Boulder, CO, 80303, USA\\
$^{11}$Space Vehicles Directorate, Air Force Research Laboratory, Kirtland AFB, NM 87117, USA
}

\subject{astrophysics, solar system}


\corres{Shahin Jafarzadeh\\
\email{shahin.jafarzadeh@astro.uio.no}}

\maketitle


\begin{tcolorbox}[sharp corners, width=\textwidth,colback=abstract-color,colframe=abstract-color,boxsep=5pt,left=0pt,right=0pt,top=0pt,bottom=0pt]
By direct measurements of the gas temperature, the Atacama Large Millimeter/sub-millimeter Array (ALMA) has yielded a new diagnostic tool to study the solar chromosphere. Here we present an overview of the brightness-temperature fluctuations from several high-quality and high-temporal-resolution (i.e., 1~and~2~sec cadence) time series of images obtained during the first two years of solar observations with ALMA, in Band~3 and Band~6, centred at around 3~mm (100~GHz) and 1.25~mm (239~GHz), respectively. The various datasets represent solar regions with different levels of magnetic flux. We perform Fast Fourier and Lomb-Scargle transforms to measure both the spatial structuring of dominant frequencies and the average global frequency distributions of the oscillations (i.e., averaged over the entire field of view). We find that the observed frequencies significantly vary from one dataset to another, which is discussed in terms of the solar regions captured by the observations (i.e., linked to their underlying magnetic topology). While the presence of enhanced power within the frequency range $3-5$~mHz is found for the most magnetically quiescent datasets, lower frequencies dominate when there is significant influence from strong underlying magnetic field concentrations (present inside and/or in the immediate vicinity of the observed field of view).
We discuss here a number of reasons which could possibly contribute to the power suppression at around 5.5~mHz in the ALMA observations. However, it remains unclear how other chromospheric diagnostics (with an exception of H$\alpha$ line-core intensity) are unaffected by similar effects, i.e., they show very pronounced 3-min oscillations dominating the dynamics of the chromosphere, whereas only a very small fraction of all the pixels in the ten ALMA data sets analysed here show peak power near 5.5~mHz.
\end{tcolorbox}


\section{Introduction}
The solar chromosphere is historically known to be dominated by 3-min ($\approx5.5$~mHz) oscillations of acoustic/magneto-acoustic waves \cite{1991A&A...250..235F,1991SoPh..134...15R,1992ApJ...397L..59C,1995ESASP.376a.151R,1997A&A...324..587T}. These oscillations have been shown to be a consequence of a resonance at the chromospheric cut-off frequency \cite{1991A&A...250..235F,1994A&A...284..976K,1995A&A...294..232S,1995A&A...294..241S}. Acoustic ($p$-mode) waves, which are excited in the upper layers of the turbulent convection zone \cite{2015LRSP...12....6K}, are found in both quiet and magnetised regions (i.e., with considerably lower magnetic flux in the former), though their amplitudes, and thus their power, are reduced in the latter \cite{1962ApJ...135..474L,1988ESASP.286..315T,1992ApJ...393..782T}. It has been shown that the power suppression (as well as its spatial extent) increases with the magnetic-field strength and/or geometric height \cite{1997ApJ...476..392H,2002A&A...387.1092J,2007A&A...471..961M,2014ApJ...796...72J,2016ApJ...823...45K}. Such power suppression in the high chromosphere, also known as ``magnetic shadows'', are thought to be due to the interaction of $p$-mode oscillations with the embedded magnetic fields (i.e., within the magnetic canopy \cite{1990A&A...234..519S,2009SSRv..144..317W}). This interaction leads to mode conversion at the acoustic-Alfv\'{e}n equipartition layer, where the sound speed equals the Alfv\'{e}n speed ($c_{s}=v_{A}$), and where the purely acoustic waves (i.e., fast mode in the high-$\beta$ regime; the plasma-$\beta$ is deﬁned as the ratio of gas and magnetic pressures) can be converted to magneto-acoustic waves (fast mode in the low-$\beta$ regime) and vice versa \cite{2007A&A...471..961M,2010AN....331..915N,2012A&A...542L..30N,2014A&A...567A..62K}. We note that a similar process can cause an opposite effect at photospheric to mid-chromospheric heights, i.e., power enhancements around magnetic-field concentrations, such as network patches (known as ``power halos''; \cite{1992ApJ...394L..65B,2010A&A...510A..41K,2016ApJ...817...45R}). Power halos have been explained as the effect of fast-mode reflection at the magnetic canopy, due to the steep Alfv\'{e}n speed gradient, resulting in power enhancements at lower chromospheric heights \cite{2012ApJ...746...68K,2016ApJ...817...45R}.

The pronounced 3-min fluctuations have often been observed in the line-of-sight velocity and/or intensity signals, in both quiet and active regions, through chromospheric spectral lines such as Ca~{\sc ii}~H \& K \cite{1969SoPh....7..351B,1994chdy.conf..103F,2008A&A...480..515C,2009A&A...500.1239R}, Ca~{\sc ii} infrared triplet \cite{1982ApJ...253..386L,1989A&A...224..245F,1990A&A...228..506D,2009A&A...500.1239R}, H$\alpha$ \cite{2017Ap&SS.362...46Z}, He~{\sc i}~1083~nm \cite{1986ApJ...301.1005L,1995itsa.conf..437F,1994chdy.conf..103F}, and Mg~{\sc ii}~h \& k \cite{1987SoPh..108...61G}. In the quiet Sun, while the 3-min oscillations have been shown in many observations of the low-to-mid chromosphere (or using relatively broad-band filters), a lack of such fluctuations in intensity image sequences of H$\alpha$ line-core observations have also been reported \cite{2016ApJ...828...23S}. \citet{2016ApJ...828...23S} showed, however, that the dominating 3-min oscillations could be detected in the line-of-sight velocities of their H$\alpha$ observations (i.e., also at the same spectral line position; the H$\alpha$ line-core), but not in the intensity. In the latter, they found the dominant periods (in the entire field-of-view) to be longer than five minutes. These authors found that the presence of ubiquitous transient events in the chromosphere were contributing in the power suppression at around 3-min periodicities, particularly, in the magnetised regions, since the intensity fluctuations were mostly due to the appearance and disappearance of such events at longer periods (i.e., 5-9 minutes). Yet, the mode conversion found to play a key role (over the entire field-of-view) in the formation of magnetic shadows at periodicities around three minutes.

Moreover, longer-period internal-gravity waves, as well as various types of magneto-hydrodynamic (MHD) waves of different periods and properties, have been observed throughout the chromosphere at all spatial scales, e.g.,~\cite{2008ApJ...681L.125S,2012NatCo...3.1315M,2014A&A...566A..90M,2017ApJS..229...10J,2017ApJS..229....9J,2017A&A...607A..46M,2020NatAs...4..220J}. Such magneto-acoustic-gravity (MAG) waves are often considered as a prime means of transporting energy through the solar atmosphere, thus, contributing to the excess heating of the solar chromosphere and beyond, where they release their energy \cite{2015SSRv..190..103J}. Furthermore, periods of the propagating MAG waves in the solar chromosphere have been shown to be dependent on the inclination of the magnetic fields (i.e., the cut-off period decreases with the field inclination \cite{1977A&A....55..239B,2006ApJ...648L.151J,2011A&A...534A..65S}). Hence, the highly inclined magnetic fields play a role in the propagation of long-period waves \cite{2013ApJ...779..168J}. Thus, distributions of the observed periods at various chromospheric heights are expected to, on one hand, depend on the amount of the magnetic flux within the field of view (FOV) under study. On the other hand, at chromospheric heights (and beyond), the distributions may be influenced by the intrusion of highly inclined magnetic fields originating in neighbouring field-concentrations/active-regions (the canopy effect; \cite{1982SoPh...79..247J,1990A&A...234..519S}). It has been shown that the height of the magnetic canopy depends on the strength of its underlying magnetic-field in the solar photosphere \cite{2017ApJS..229...11J}, ranging from the middle photosphere to the high chromosphere and beyond. In addition, interactions between the magnetic fields at various chromospheric heights (resulting in, e.g., magnetic-reconnection) may also cause oscillations and instabilities \cite{2016ApJ...819L..24Y}.

Most of the chromospheric diagnostic discussed above suffer from non-local thermodynamic equilibrium (non-LTE) formations, thus they are decoupled from the local conditions \cite{2019ARA&A..57..189C}. Alternatively, the continuum emissions at millimetre wavelengths where the Rayleigh–Jeans law holds, are formed under LTE conditions that can be linearly correlated the gas temperatures at various chromospheric heights (i.e., the recorded brightness temperatures are equivalent to the gas temperatures).

In December 2016, the Atacama Large Millimeter/sub-millimeter Array (ALMA; \cite{2009IEEEP..97.1463W}) started regular observations of the solar chromosphere at millimetre wavelengths, providing new capabilities (for studying, e.g., waves and oscillations), namely, direct measurements of the gas temperature and stable observations at high temporal resolutions (i.e., $1-2$~sec cadences). These particular aspects of ALMA observations thus represent a unique opportunity for the study of heating mechanisms associated to waves. ALMA is a millimetre/submillimetre interferometer, located at an elevation of about 5~km above sea level in the Atacama desert in northern Chile, providing excellent atmospheric transmission over the wavelength range $0.3-8$~mm \cite{1999PASJ...51..603M}. The solar radiation in this range originates from the chromosphere
\cite{2002AN....323..271B,2016Msngr.163...15W,2017SoPh..292...88W,2017A&A...601A..43L,2018Msngr.171...25B}. ALMA is currently (i.e., September 2020) in its fourth year of solar science operations. Thus far, Band~3 (centred near 3~mm) and Band~6 (centred at $\approx1.25$~mm), with 1~sec and 2~sec sampling cadences, have been provided for observing the Sun, both of which supposedly sample the mid-to-high chromosphere \cite{2016SSRv..200....1W,2017ApJ...845L..19B,2018A&A...619L...6N,2019ApJ...877L..26L,2020A&A...635A..71W}. Their precise formation heights are, however, not known to date and are predicted (from numerical simulations and solar models) to span a large range between the low chromosphere and the transition region \cite{2020ApJ...891L...8M}. Numerical simulations have predicted the importance of millimetre observations in identifying upper chromospheric dynamics \cite{2004A&A...419..747L,2007A&A...471..977W,2015A&A...575A..15L} and in the detection of 3-min chromospheric oscillations at those wavelengths with, e.g., ALMA \cite{2006A&A...456..713L,2010MmSAI..81..592L}. However, \citet{2017A&A...598A..89R} speculated that such oscillations would be hidden due to the presence of the magnetic canopy and as a result, an increase in opacity in the millimetre wavelengths at mid-to-high chromospheric heights, acting as an ``umbrella'' obscuring the dynamics underneath (a similar effect to that seen in H$\alpha$ observations). \citet{2017A&A...598A..89R} showed that fibrillar structures, which are thought to highlight the magnetic canopy in intensity images, e.g., in H$\alpha$ imaging observations, may not be clearly seen in ALMA observations due to their reduced lateral contrast (due to an insensitivity to Doppler shifts) at those wavelengths. In \citet{2019ApJ...881...99M}, it is pointed out that fibrillar structures, similar to some of those visible in H$\alpha$ line width, are also present in the ALMA 3~mm brightness temperature, down to the 2~arcsec resolution of ALMA.

This paper presents an overview of the properties of (global) chromospheric fluctuations observed at millimetre wavelengths with ALMA. The analysed datasets (across various solar regions with different levels of magnetic flux) and their calibrations are briefly described in Section~\ref{section_data}. Section~\ref{section_analysis} summarises the approaches with which the power spectra are calculated, as well as, the results for the several employed datasets. The discussion and concluding remarks are presented in Section~\ref{section_conclusion}.

\section{Data}
\label{section_data}

We utilise ten different datasets in this article, acquired during the first two years of solar observations with ALMA in Band~3 and 6, centred at around 3~mm (100~GHz) and 1.25~mm (239~GHz), respectively. The time of observation for each dataset, the project's identifier number, cadence of the observations, the cosine of the heliocentric angle ($\mu$), and the mean temperature (averaged over the entire time series) are summarised in Table~\ref{table_obslog}.

\vspace*{-6pt}
\begin{table}[!h]
\caption{Summary of the ALMA datasets employed in this study.}
\label{table_obslog}
\setlength{\tabcolsep}{.53em}
\begin{tabular}{lccccccc}
\hline
Data & Date & Project ID & Band & Cad. [sec] & Obs. Time (UTC) & $\mu$ & $T_{mean}$ [K] \\
\hline
D1 & 2016-12-22 & 2016.1.00423.S & 3 & 2 & 14:19:31-15:07:07 & 0.99 & $7470\pm480$\\
D2 & 2017-04-22 & 2016.1.00050.S & 3 & 2 & 17:20:25-17:54:54 & 0.92 & $9465\pm1129$\\
D3 & 2017-04-23 & 2016.1.01129.S & 3 & 2 & 17:19:19-18:52:54 & 0.96 & $7026\pm1564$\\
D4 & 2017-04-27 & 2016.1.01532.S & 3 & 2 & 14:19:52-15:31:17 & 0.78 & $7728\pm1150$\\
D5 & 2017-04-27 & 2016.1.00202.S & 3 & 2 & 16:00:30-16:43:56 & 0.96 & $7420\pm1333$\\
D6 & 2018-04-12 & 2017.1.00653.S & 3 & 1 & 15:52:28-16:24:41 & 0.90 & $7663\pm625$\\
\hline
D7 & 2017-04-18 & 2016.1.01129.S & 6 & 2 & 14:22:01-15:09:15 & 0.76 & $7324\pm954$\\
D8 & 2017-04-22 & 2016.1.00050.S & 6 & 2 & 15:59:07-16:43:26 & 0.92 & $7746\pm859$\\
D9 & 2018-04-12 & 2017.1.00653.S & 6 & 1 & 13:58:58-14:32:27 & 0.88 & $5957\pm352$\\
D10 & 2018-08-23 & 2017.1.01672.S & 6 & 1 & 16:24:27-17:18:05 & 0.97 & $6104\pm497$\\
\hline
\end{tabular}
\vspace*{-4pt}
\end{table}

The datasets were acquired using various array configurations of the 12~m and 7~m antennas (depending on the observing programme/cycle). Each ALMA band consists of four so-called spectral windows (or sub-bands) with 128 spectral channels each. In this study, we use the band-averaged images which have the highest signal-to-noise ratio (S/N), compared to individual channels or sub-band-averaged images. 

Although there is a small height difference between the sub-bands \cite{2019A&A...622A.150J}, the high S/N is of great importance in the present study to identify potentially small-amplitude oscillations. In addition to the interferometric images, full-disc total power (TP) maps were also recorded for each dataset through single-dish observations (i.e. with the TP array in fast-scanning mode). These were used to calibrate absolute brightness temperatures for the interferometric datasets \cite{2017SoPh..292...87S}. The mean brightness temperatures were then corrected based on the values provided by \citet{2017SoPh..292...88W}. All procedures were employed through the Solar ALMA Pipeline (SoAP; Szydlarski et al., in prep.). For further details on the reduction procedures we refer the reader to \citet{2020A&A...635A..71W}.
We note that the time series of images were recorded in several $8-10$~min blocks, with $\sim1.5-3$~min gaps in between for calibrations. Thus, these gaps must be taken into account when an entire image sequence is analysed (see Section~\ref{section_analysis}). 

Furthermore, we make use of full-Stokes observations (of Fe~{\sc i}~617.3~nm) from the Helioseismic and Magnetic Imager (HMI; \cite{2012SoPh..275..229S}) on board the Solar Dynamics Observatory (SDO; \cite{2012SoPh..275....3P}), in order to obtain the magnetic-field topology of the field-of-view (FOV) of the ALMA datasets. The SDO images were spatially co-aligned with the corresponding ALMA images, at a selected time during each image sequence. We use a combination of the 170~nm and 30.4~nm channels (with different intensity weights) from the SDO's Atmospheric Imaging Assembly (AIA; \cite{2012SoPh..275...17L}) to perform a precise spatial co-alignment with the ALMA images. The combined AIA 170/30.4~nm image resulted in a similar scene to that observed in ALMA’s Band~3 and/or Band~6, hence, facilitated the cross correlation of similar solar features in the alignment procedure. HMI's full-Stokes parameters were then inverted with the Milne–Eddington's VFISV code \cite{2011SoPh..273..267B} to infer the full-vector photospheric magnetic fields. These are further used for non-force-free-field and potential-field extrapolations (see Section~\ref{section_analysis}). Table~\ref{table_hmi} summarises the photospheric magnetic-field properties, as well as solar $x$ and $y$ coordinates of the centre of the observed FOVs, of the ten datasets studied here. Magnetic flux and B$_{los}$ were calculated for both the ALMA's FOVs and extended regions (which are larger by a factor of two compared to those of ALMA; see bottom-left panels in Figures~\ref{fig_D1}-\ref{fig_D10}). The latter can provide information about the immediate vicinity of the observed regions, which have influence on magnetic configurations at chromospheric heights sampled by the ALMA observations. The magnetic fluxes within the extended FOVs of D6 and D9 datasets are considerably (i.e., by $1-3$ orders of magnitude) smaller compared to the other datasets (D5 has also a relatively low magnetic flux).

\vspace*{-6pt}
\begin{table}[!h]
\caption{Photospheric magnetic-field properties of the ALMA datasets (see Table~\ref{table_obslog}) from SDO/HMI.}
\label{table_hmi}
\setlength{\tabcolsep}{.54em}
\begin{tabular}{lcccccc}
\hline
Data & Solar Coordinates & \multicolumn{2}{c}{ALMA's FOV} && \multicolumn{2}{c}{Extended FOV}\\
\cline{3-4} \cline{6-7}
  & of centre of FOV & Magnetic Flux & Range of B$_{los}$ && Magnetic Flux & Range of B$_{los}$\\
  & ($x$,$y$) [arcsec] & [$10^{18}$~Mx] & (min,max) [G] && [$10^{18}$~Mx] & (min,max) [G]\\
\hline
D1 & $(6,-2)$ & \num{1.58e1} & $(-284,276)$ && \num{3.32e2} & $(-1164,634)$ \\
D2 & $(-246,267)$ & \num{1.46e3} & $(-1084,59)$ && \num{2.50e3} & $(-1084,534)$ \\
D3 & $(-54,251)$ & \num{2.72e2} & $(-488,65)$ && \num{3.28e3} & $(-2343,216)$ \\
D4 & $(520,276)$ & \num{1.12e3} & $(-823,1240)$ && \num{1.72e3} & $(-913,1240)$ \\
D5 & $(172,-207)$ & \num{1.11e2} & $(-82,629)$ && \num{1.23e2} & $(-138,629)$ \\
D6 & $(-128,400)$ & \num{5.5e-1} & $(-78,110)$ && \num{2.42e1} & $(-225,141)$ \\
\hline
D7 & $(-573,230)$ & \num{1.38e2} & $(-37,450)$ && \num{1.80e2} & $(-378,450)$ \\
D8 & $(-255,263)$ & \num{5.01e2} & $(-949,48)$ && \num{1.46e3} & $(-1116,81)$ \\
D9 & $(-175,-415)$ & \num{6.57e0} & $(-25,132)$ && \num{9.96e0} & $(-63,132)$ \\
D10 & $(68,-211)$ & \num{2.45e0} & $(-1066,1069)$ && \num{4.74e2} & $(-1261,1977)$ \\
\hline
\end{tabular}
\vspace*{-4pt}
\end{table}

\section{Analysis and Results}
\label{section_analysis}

To investigate the presence of oscillatory phenomena in our ALMA datasets, we perform Fourier analysis on individual pixels from the entire FOV of each dataset. The time series were de-trended (by subtracting a simple linear fit) and apodised (by using the Tukey window), prior to the analysis. As a result, power spectra are calculated at each pixel.
Here, we are primarily interested in identifying the global oscillatory behaviours of the observations, therefore, an averaged power spectrum over the entire FOV is also calculated (i.e., we take the average of all individual power spectra from all pixels within the FOV of each dataset). 

In order to compute the power spectra, we employ two different Fourier methods, namely, the classical Fast Fourier Transform (FFT) and the Lomb-Scargle approach \cite{1976Ap&SS..39..447L,1982ApJ...263..835S}. While the former method is often considered as the prime choice for such analyses, the Lomb-Scargle transform is a well-known statistical technique for identifying oscillation frequencies in irregularly-sampled signals. Thus, the Lomb-Scargle approach (which estimates a frequency spectrum based on a least squares fit of sinusoids) allows us to accurately measure the oscillations present in the ALMA observations and is unaffected by the data gaps (which are due to ALMA's routine calibration requirements). Yet, for comparison (and for the power spectra $k-\omega$ diagrams), we also employ the FFT, for which some approximations are necessary (i.e., the FFT is based on the assumption that all samples are evenly spaced with time). Thus, the gaps are filled in by means of linear (spline) interpolation into a time series with uniform temporal sampling prior to the FFT, whereas the original time series (which include gaps/missing frames) are used for the Lomb-Scargle transforms. The $k-\omega$ diagrams represent the average (FFT) power spectra in the wavenumber-frequency domain, where the distribution of power is shown in both wavenumber (or spatial scale) and frequency (or period) space. In addition, we calculate the peak-frequency maps (i.e., frequencies corresponding to the maximum power at all individual pixels) to investigate their spatial distributions over the entire FOV. We note that such peak-frequency maps should be interpreted with caution since they may suffer from an uncertainty at places where the power spectra pose multiple peaks (at different frequencies) with small power differences. Hence, the frequency of the absolute maximum power may not necessarily identify the significant frequency at a pixel, but it can still be a representative of the major oscillation's frequency.

Furthermore, we aim to compare the spatial distribution of the peak-frequency maps
(and the observed average power spectra) with the magnetic topology of the mid-to-upper chromosphere, for which we perform magnetohydrostatic (MHS) and potential-field extrapolations of the surface magnetic fields from SDO/HMI. The MHS non-force-free field extrapolations which take into account plasma forces \cite{2018ApJ...866..130Z,2019A&A...631A.162Z,2020A&A...640A.103Z} can more reliably return the magnetic topology within the mixed plasma-$\beta$ in the upper photosphere and chromosphere, compared to the traditional force-free extrapolations (which are limited to the low plasma-$\beta$ in the solar corona \cite{2004SoPh..219...87W,2005A&A...433..701W}). Deviations from the non-force-free fields occur above $\approx2$~Mm. However, for the most magnetically quiescent datasets (i.e., D5, D6, and D9), with the small amount of magnetic flux in their extended FOVs, the MHS model could not be applied reliably. Thus, magnetic configurations for these datasets were approximated using the potential-field extrapolations, as a special
case with $\alpha=0$ of a linear force-free field using a Fast-Fourier-Transform \cite{1981A&A...100..197A}. The extrapolations are performed in considerably larger FOVs, compared to those of ALMA, and are extended to much larger geometric heights (i.e. up to $\approx14$~Mm), though we are primarily interested in the field topology within the ALMA's FOVs and up to the geometric heights corresponding to the mid-to-high chromosphere (i.e., 1500-2500~km). The larger extents reduce possible effects of the sides and top boundaries.

We note that due to uncertainty in the exact heights of formation of the ALMA observations, the comparison of the magnetic topology of the mid-to-high chromosphere and the oscillatory behaviours in the ALMA datasets are performed qualitatively. In addition, due to the fact that the magnetic configuration in the high chromosphere is largely affected by the inclined/horizontal fields from neighbouring field concentrations/active regions, no quantitative separations in each dataset between, e.g., network and internetwork regions, are provided.

In the following, we present each dataset, along with their oscillatory and magnetic-topology properties, in separate figures (i.e., in Figures~\ref{fig_D1}-\ref{fig_D10}). Each figure consists six panels organised as,
\begin{itemize}[leftmargin=0.7cm]
  \item  {\bf{Upper left:}} a brightness temperature map (sampled by the ALMA Band~3 or Band~6) of each dataset, as noted in the figure's caption. In the lower-left corner of each brightness-temperature map, the shape and angle of ALMA's (synthesised) beam has been depicted. The beam is a representative of point spread function (PSF) of the interferometric array, i.e., an elliptical Gaussian whose size and orientation angle depend on the maximum baseline of the array and the angle of the Sun with respect to the north celestial pole (i.e., the `position angle'), respectively. Thus, the beam size is a measure of the spatial resolution and its shape varies during the day, though its variation within, e.g., one hour is very small (see \citet{2020A&A...635A..71W} for further details). Thus, it is more elongated when the observations are made away from the local noon. This may have some visual effects on the structures seen in the images (in the spatial domain), so should be kept in mind, otherwise, it has no influence on the oscillatory properties.
  \item {\bf{Upper right:}} the $k-\omega$ power spectrum, using FFT with the data gaps interpolated.
  \item {\bf{Middle left:}} the spatially averaged brightness-temperature power spectra from FFT (dash-dotted black line) and Lomb-Scargle (solid red line) approaches. The FFT power spectra were calculated from the gaps-interpolated dataset, with its original length, whereas, the Lomb-Scargle power spectra were obtained from the original data (with gaps), padded with zeros (at one end) to increase the length of the time series by a factor of 4. The padding increases the frequency resolution.
  \item {\bf{Middle right:}} the peak-frequency map (from Lomb-Scargle transform), for frequencies larger than 1~mHz. The latter criterion is to focus on dominant frequencies which are important in terms of wave propagation throughout the solar chromosphere, rather than the slow variations which could be ascribed to the intrinsic evolution of the magnetic fields. To facilitate the comparison of these maps across all datasets, each peak-frequency map is plotted using the same frequency range of $1-7$~mHz (i.e., the most common range in all datasets).
  \item {\bf{Bottom left:}} the SDO/HMI magnetogram (i.e., the line-of-sight component of the photospheric magnetic fields) acquired at the same time as the brightness-temperature image shown in the upper-left panel. For clarity, this magnetogram is shown for a FOV which is larger (by a factor of 2) compared to that of ALMA; the ALMA's FOV is marked with the dashed square.
  \item {\bf{Bottom right:}} top view of the field-extrapolated cube at chromospheric heights (for ALMA’s FOV). The colours indicate the magnetic-field inclination angles, from vertical (blue) to horizontal (red). The vertical scales are quintupled for better visibility.
\end{itemize}

Each dataset (figure) is described separately in the following subsections. Possible interpretations/speculations of these observations are discussed in Section~\ref{section_conclusion}.

\begin{figure}[!htp]
\vspace*{-30pt}
\centering\includegraphics[width=\textwidth]{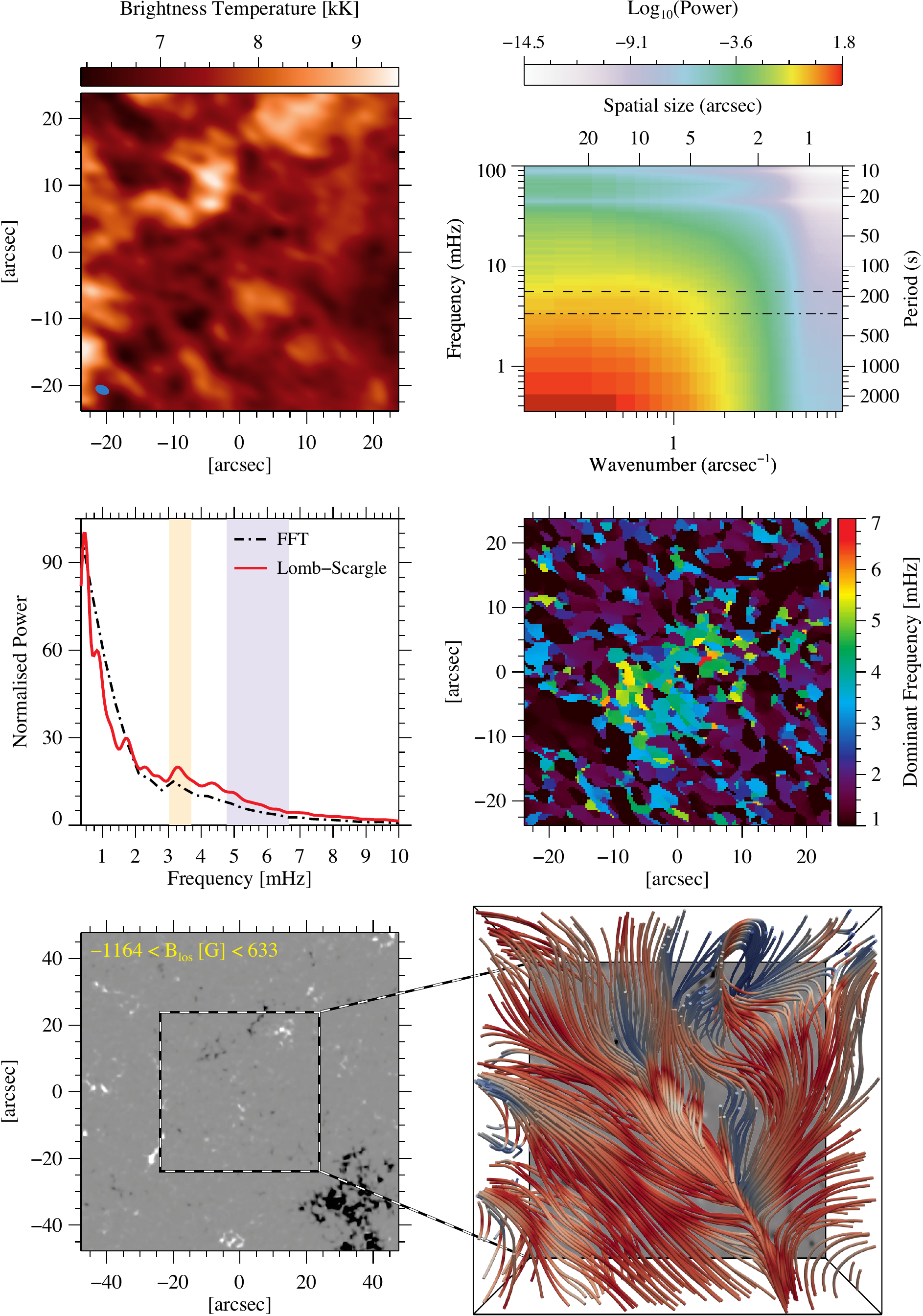}
\caption{
\textit{\textbf{Top left}}: A brightness-temperature map from the ALMA's Band-3 observations from 2016 December 22$^{\text{nd}}$ (i.e., D1 dataset). The blue ellipse on the bottom-left corner of the panel represents the beam size of the observations.
\textit{\textbf{Top right}}: $k-\omega$ power spectra from the Fourier transform. The dashed and dotted-dashed lines mark the 3- and 5-min periodicities, respectively.
\textit{\textbf{Middle left}}: Spatially averaged brightness-temperature power spectra from Fast Fourier (dash-dotted black line) and Lomb-Scargle (solid red line) transforms. The purple and yellow stripes have been depicted to mark period ranges corresponding to the 3 and 5~min windows (each with a width of 1~min), respectively.
\textit{\textbf{Middle right}}: Dominant-frequency map (i.e., frequencies corresponding to the largest power at each pixel) from the Lomb-Scargle approach, for frequencies longer than 1~mHz.
\textit{\textbf{Bottom left}}: Line-of-sight photospheric magnetic fields (B$_{los}$) from SDO/HMI with a factor of two larger field-of-view than that of ALMA. The range of B$_{los}$ values has been indicated in the upper left corner. The ALMA's field-of-view is marked with the dashed square. 
\textit{\textbf{Bottom right}}: Top view of field extrapolation of the surface magnetic field (from SDO/HMI) at the chromosphere heights (for the ALMA's field-of-view). The colours represent inclination, from vertical (blue) to horizontal (red). The vertical scales are quintupled for better visibility.
}
\label{fig_D1}
\end{figure}

\subsection{D1 dataset: ALMA Band 3, 2016 December 22$^{\text{nd}}$ (Figure~\ref{fig_D1})}

The dataset (also appeared in \citet{2020A&A...635A..71W} and \citet{Eklund2020RS}) samples a relatively quiet area at solar disc centre, with a few small network patches of opposite polarities, mostly located towards the top of the FOV, but also with some weaker magnetic patches located close to the centre of the FOV (as seen in the HMI magnetogram). These are observed as excess brightness temperatures in ALMA's Band~3 images. However, there are large field concentrations in the immediate vicinity of the ALMA's FOV, e.g., close to the bottom-right corner. Thus, the ALMA's FOV is largely influenced by the horizontal fields (i.e., the magnetic canopy) rooted in the neighbouring active regions. The magnetic topology seems to have some complexities at a few locations within the FOV, which is likely due to the interaction of the horizontal fields with the more vertical ones originating from the small network patches within the FOV.

The $k-\omega$ diagram and the mean power spectra show a clear lack of prominent chromospheric 3-min oscillations (cf. see Figure~2 of \citet{2017ApJ...842...59J} for a clear power enhancement at 3~min). Most of the power is concentrated at very low frequencies (lower than 1~mHz) and at relatively large spatial scales (i.e., larger than 14 arcsec), with only very small enhancements at various higher frequencies (which are not significant). While such low frequencies dominate the entire FOV, there also exist higher frequencies, up to 7~mHz, within parts of the FOV, mostly around the centre with frequencies in the range of $3-7$~mHz. No clear relationship between their spatial locations and the chromospheric magnetic topology is found.

\begin{figure}[!htp]
\centering\includegraphics[width=\textwidth]{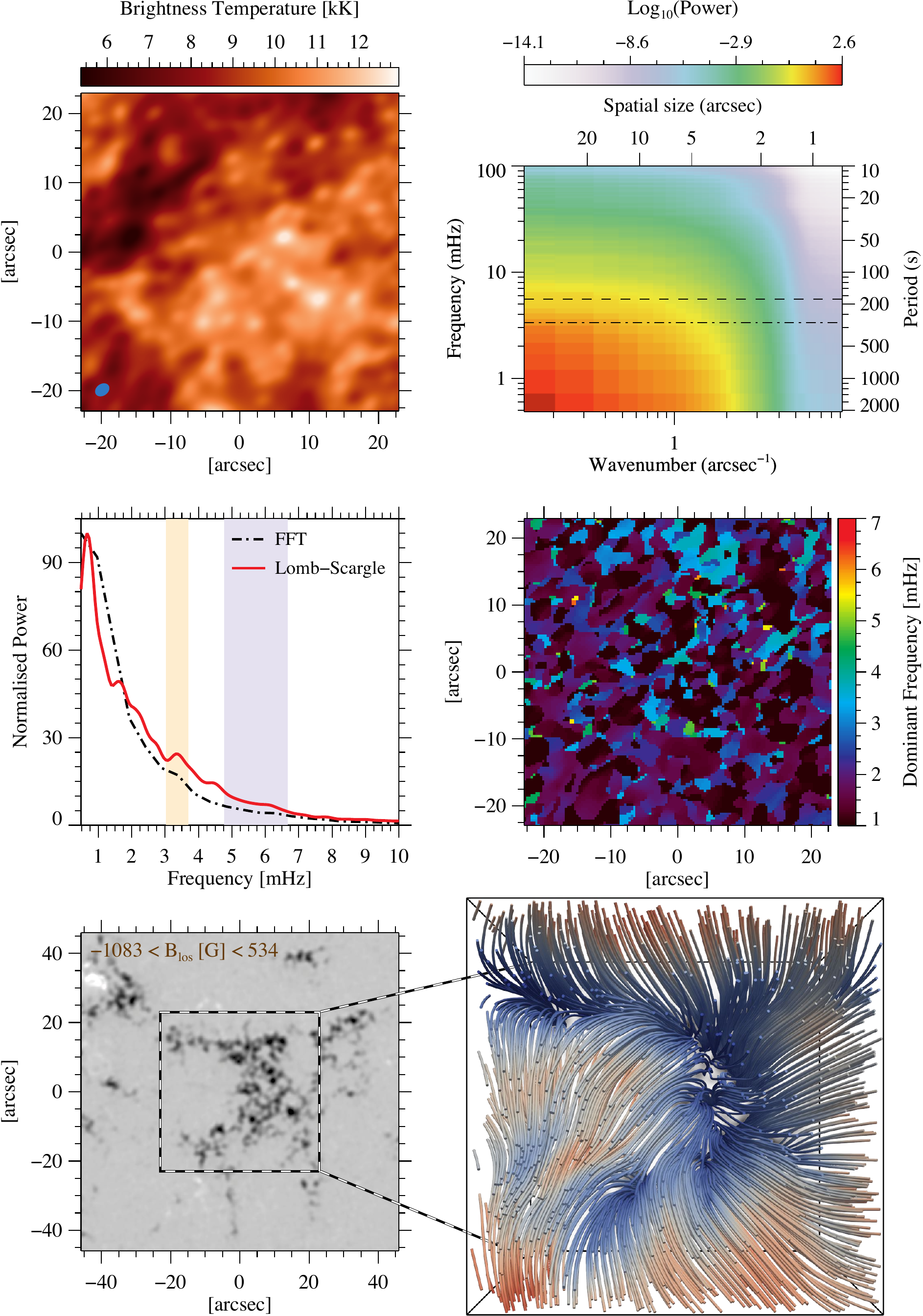}
\caption{
Same as Figure~\ref{fig_D1}, but for ALMA's Band-3 observations from 2017 April 22$^{\text{nd}}$ (i.e., D2 dataset).
}
\label{fig_D2}
\end{figure}

\subsection{D2 dataset: ALMA Band 3, 2017 April 22$^{\text{nd}}$ (Figure~\ref{fig_D2})}

This observation (which is also used in \citet{Guevara-Gomez2020}) samples a plage region, with strong magnetic-field concentrations covering the majority of the FOV (organised in an inverse c-shape) in the photosphere. As a result, the magnetic topology at chromospheric heights largely represents the vertical/near-vertical fields, but also more inclined/horizontal fields towards the edges of the FOV and within the areas between the centre and left edge of the FOV. A clear lack of dominant frequencies larger than 2~mHz is observed over the entire FOV (i.e., much of the power is concentrated in frequencies smaller than $1-2$~mHz and in spatial scales larger than 31~arcsec). However, there are small areas with dominant frequencies around 3~mHz that seem to coincide with more vertical-field regions.

\begin{figure}[!htp]
\centering\includegraphics[width=\textwidth]{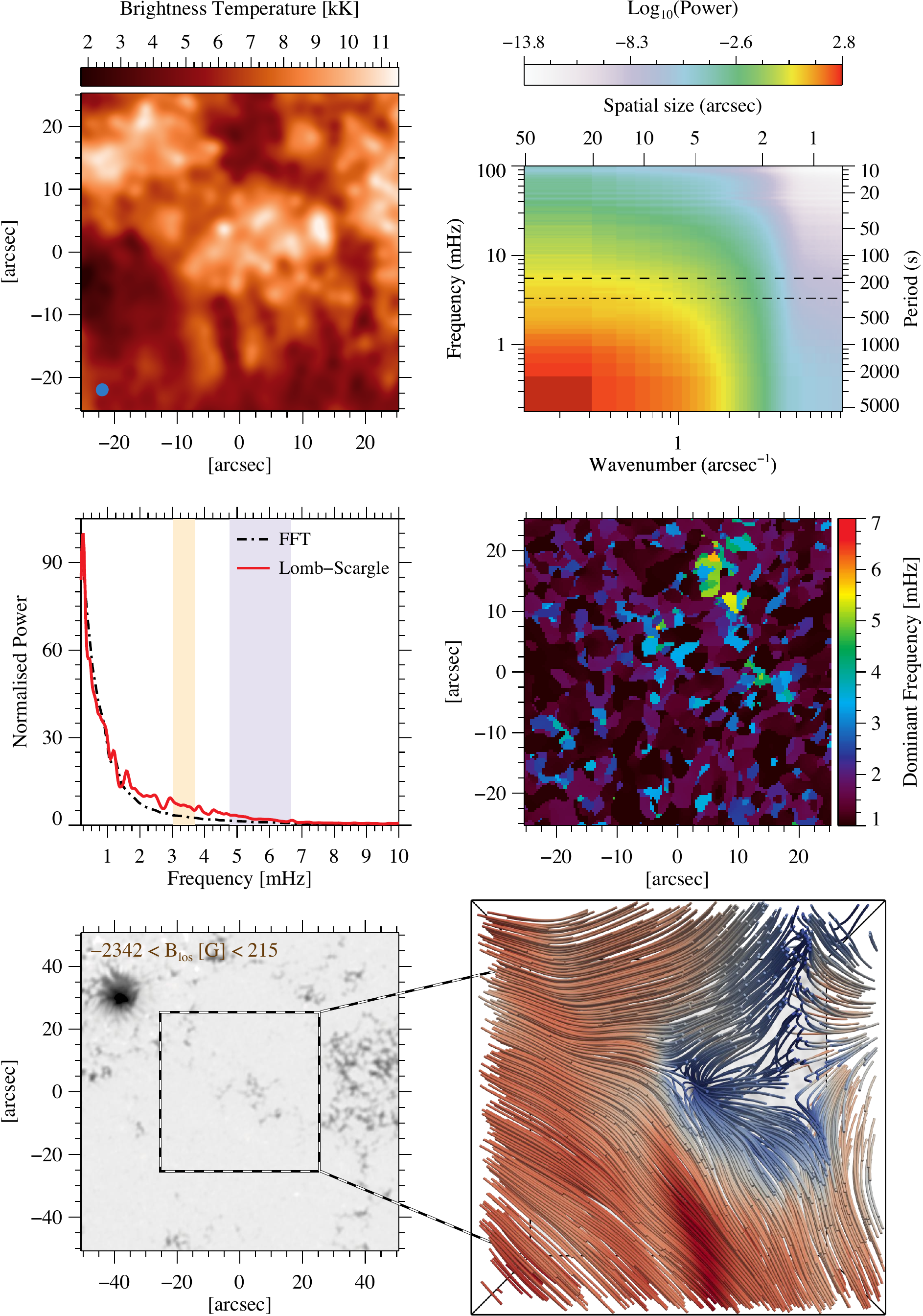}
\caption{
Same as Figure~\ref{fig_D1}, but for ALMA's Band-3 observations from 2017 April 23$^{\text{rd}}$ (i.e., D3 dataset).
}
\label{fig_D3}
\end{figure}

\vspace*{-3pt}
\subsection{D3 dataset: ALMA Band 3, 2017 April 23$^{\text{rd}}$ (Figure~\ref{fig_D3})}

The observation (which was also appeared in \citet{2019ApJ...881...99M}) was made in the immediate vicinity of a sunspot and includes a few (strong) magnetic concentrations within the FOV, mostly between the centre and right edge.  The magnetic canopies from the sunspot and neighbouring plage regions cover the majority of ALMA's FOV in the chromosphere with horizontal fields, although there is an exception around the strong plage patch where some near-vertical fields have changed the magnetic structuring. Only around this area some higher frequencies in the range of $3-6$~mHz can be observed in the peak-frequency map. However, the power spectra indicate dominant low frequencies (<1~mHz), mostly on spatial scales on the order of 20~arcsec or larger.

\begin{figure}[!htp]
\centering\includegraphics[width=\textwidth]{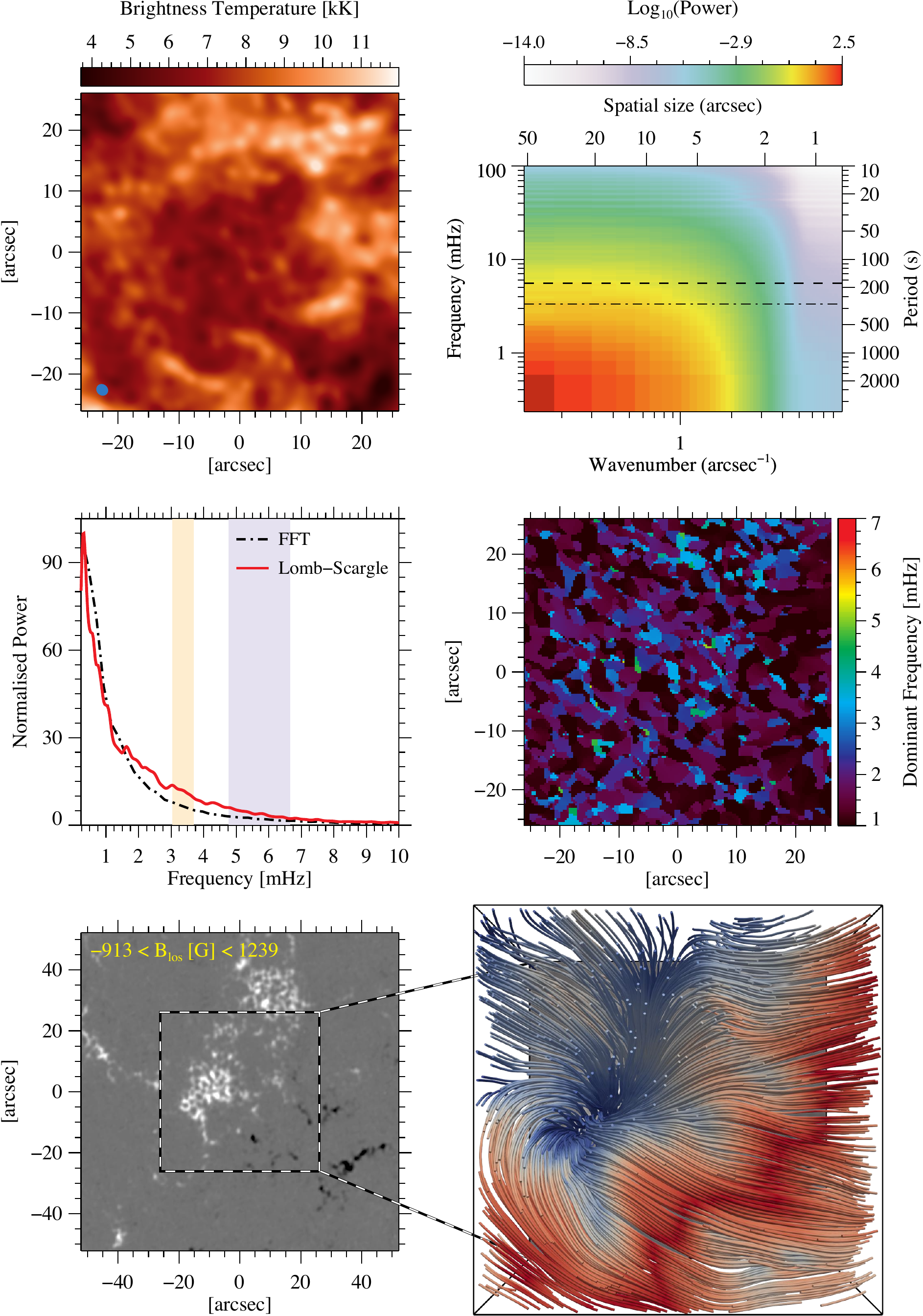}
\caption{
Same as Figure~\ref{fig_D1}, but for ALMA's Band-3 observations from 2017 April 27$^{\text{th}}$ (i.e., D4 dataset).
}
\label{fig_D4}
\end{figure}

\vspace*{-3pt}
\subsection{D4 dataset: ALMA Band 3, 2017 April 27$^{\text{th}}$ (Figure~\ref{fig_D4})}

This dataset represents large enhanced-network patches (of the same polarity) close to the centre of ALMA's FOV, towards the left edge and the upper-left corner. This has, therefore, created vertical/near-vertical fields at chromospheric heights in those photospheric network locations, and more horizontal fields (i.e., the magnetic canopy) towards the right and bottom and over the internetwork areas. No sign of significant high frequencies (higher than $1-2$~mHz) is found. There are higher frequencies, up to $\approx5$~mHz in a few pixels in various (random) locations, but these are in a minority over the entire FOV. Most of the low frequencies are also concentrated in larger structures, with the peaks of the power spectra corresponding to spatial scales larger than 40~arcsec (see the $k-\omega$ diagram).

\begin{figure}[!htp]
\centering\includegraphics[width=\textwidth]{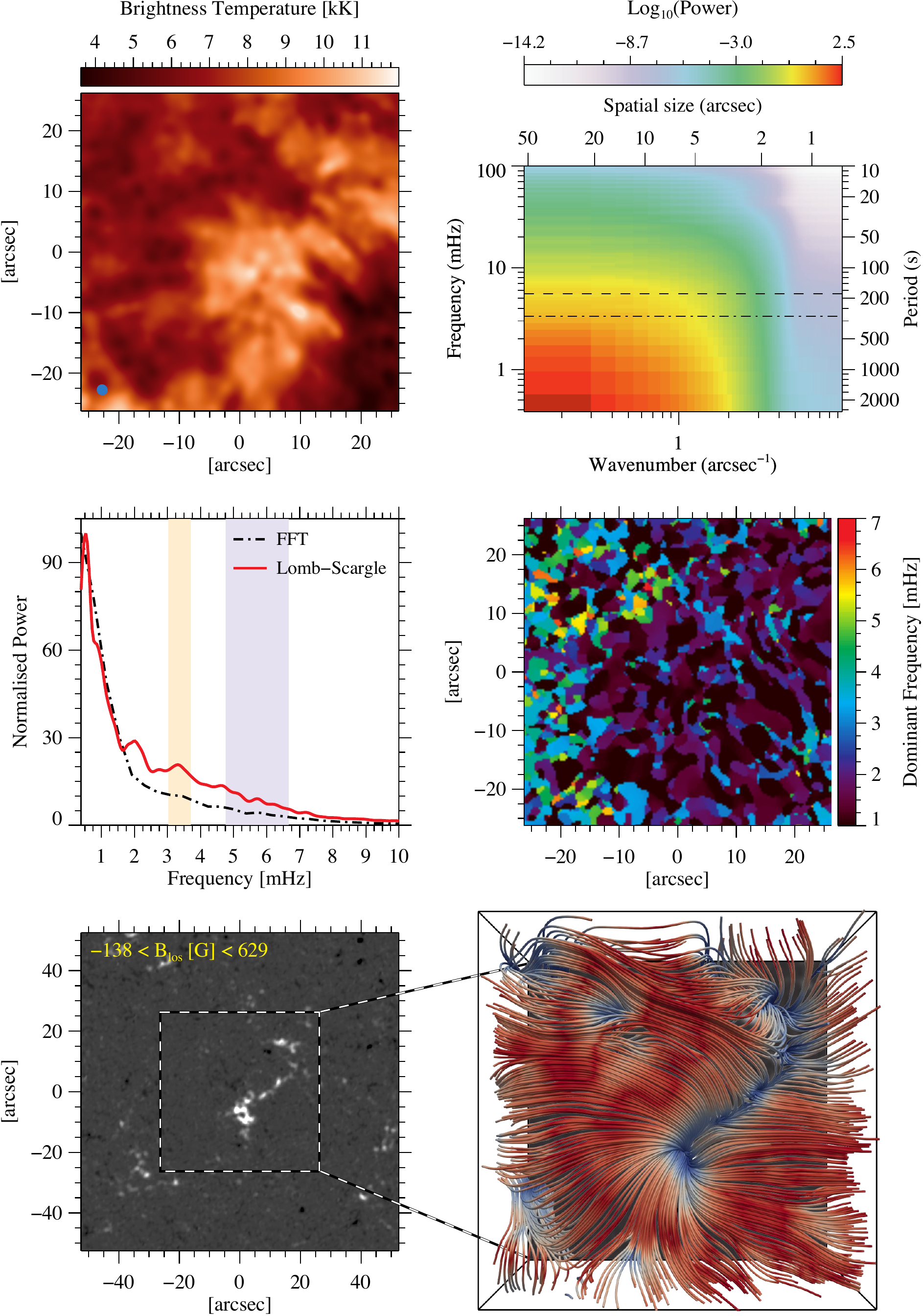}
\caption{
Same as Figure~\ref{fig_D1}, but for ALMA's Band-3 observations from 2017 April 27$^{\text{th}}$ (i.e., D5 dataset).
}
\label{fig_D5}
\end{figure}

\vspace*{-3pt}
\subsection{D5 dataset: ALMA Band 3, 2017 April 27$^{\text{th}}$ (Figure~\ref{fig_D5})}

The observations (previously appeared in \citet{2019ApJ...877L..26L} and \citet{2020ApJ...891L...8M}) were made in a relatively quiet region, where a tilted question-mark-shaped network patch is observed in the photospheric magnetogram (towards the right edge of ALMA's FOV). The chromospheric fields are closer to vertical in the spatial locations above the photospheric network patches, while the horizontally oriented magnetic canopy, mostly rooted in the network patch, covers the remainder of the FOV. Some of the magnetic fields return to the solar surface, which are visible towards the left edge, particularly at the upper- and lower-left corners of the ALMA's FOV. Although the power spectra show much of their power at frequencies lower than 2~mHz (mostly on spatial scales larger than 21~arcsec), there are also some small power enhancements at around $3-4$~mHz. The peak-frequency map also shows more concentrations of $3-6$~mHz oscillations towards the upper- and lower-left corners, i.e., where the magnetic canopy connects to the solar surface on its weaker-field side (i.e., opposite to the stronger network patch location).

\begin{figure}[!htp]
\centering\includegraphics[width=\textwidth]{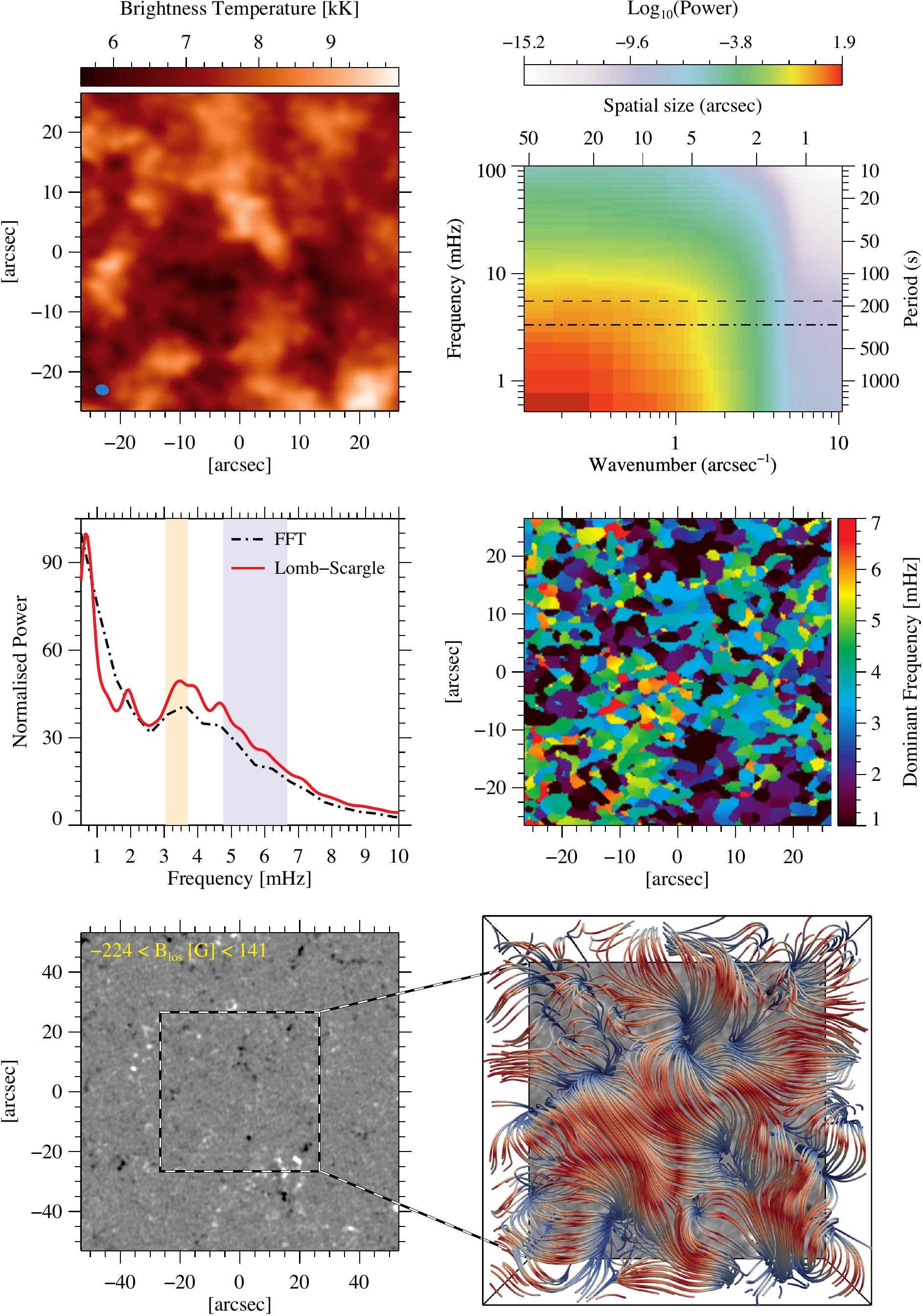}
\caption{
Same as Figure~\ref{fig_D1}, but for ALMA's Band-3 observations from 2018 April 12$^{\text{th}}$ (i.e., D6 dataset).
}
\label{fig_D6}
\end{figure}

\vspace*{-2pt}
\subsection{D6 dataset: ALMA Band 3, 2018 April 12$^{\text{th}}$ (Figure~\ref{fig_D6})}

This observation (also studied in \citet{2020arXiv200609886A}) represents the most quiescent solar region amongst our sample. ALMA's FOV (and its immediate vicinity) includes several small magnetic concentrations, which seem to form some relatively weak network and internetwork areas. These have formed some relatively low-lying magnetic canopies whose spatial extents are also relatively small, when compared to the larger chromospheric canopies originating in stronger/larger field concentrations (which reside mostly in the mid-to-high chromosphere). Although the fields from these small concentrations may also reach the higher chromosphere, they are less dense compared to those from stronger fields. More importantly, no strong/large magnetic fields were located close to the observed FOV, thus, no intrusion of horizontal fields is observed in the field geometries obtained from the extrapolations. The power spectra show a clear enhancement at around $3-5$~mHz, and the peak-frequency map reveals frequencies up to 7~mHz (also up to 9~mHz in a few pixels), mostly on top of the photospheric internetwork regions (i.e., on the relatively short and horizontally oriented field lines). At the locations of more vertical fields, lower frequencies (on the order of $1-2$~mHz) are observed.

\begin{figure}[!htp]
\centering\includegraphics[width=\textwidth]{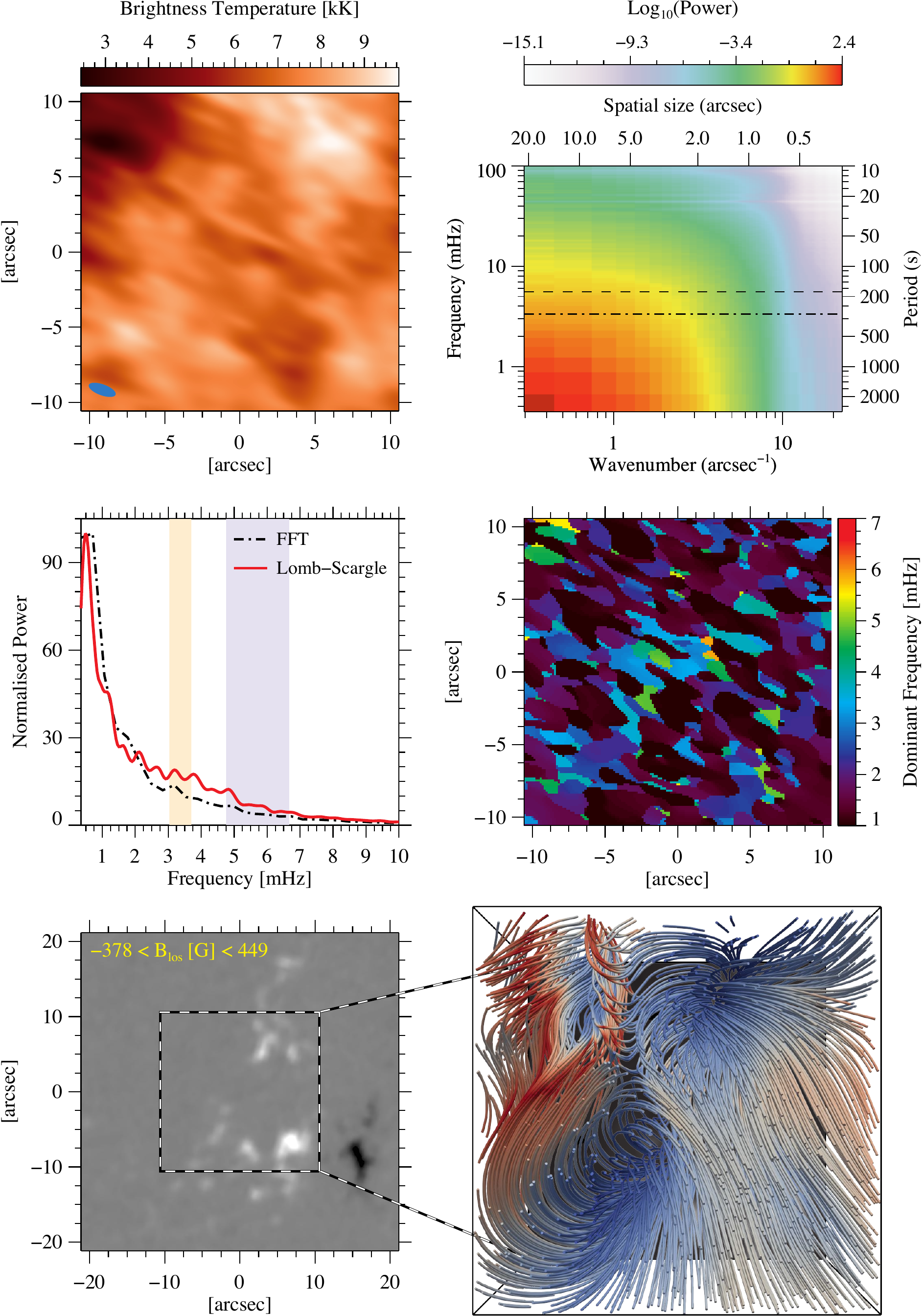}
\caption{
Same as Figure~\ref{fig_D1}, but for ALMA's Band-6 observations from 2017 April 18$^{\text{th}}$ (i.e., D7 dataset).
}
\label{fig_D7}
\end{figure}

\vspace*{-3pt}
\subsection{D7 dataset: ALMA Band 6, 2017 April 18$^{\text{th}}$ (Figure~\ref{fig_D7})}

The FOV is largely influenced by two strong magnetic regions (at the base of the photosphere), one located close to the upper-right corner, the other towards the lower half of the FOV. The field geometries in the mid-to-high chromosphere are populated by slightly inclined and near-vertical fields from the network patches within the FOV, with some horizontal fields close to the left/upper-left edge (some coming in from neighbouring magnetic concentrations). The magnetic topology is rather complex, with an inclined spiral-shape configuration near the centre of the FOV towards the lower-left corner. From the peak-frequency map, small areas with dominant frequencies of $3-5$~mHz are seen close to the root of the spiral configuration. Similar frequencies are also found in a very small area in the upper-left corner, where an intrusion of horizontal field canopy is observed (i.e., a very cool area, compared to the rest of the FOV). Otherwise, in the majority of the FOV (as also seen from the power spectra), much of the power is in the very low-frequency regime (also mostly concentrated in spatial scales larger than 14~arcsec).

\begin{figure}[!htp]
\centering\includegraphics[width=\textwidth]{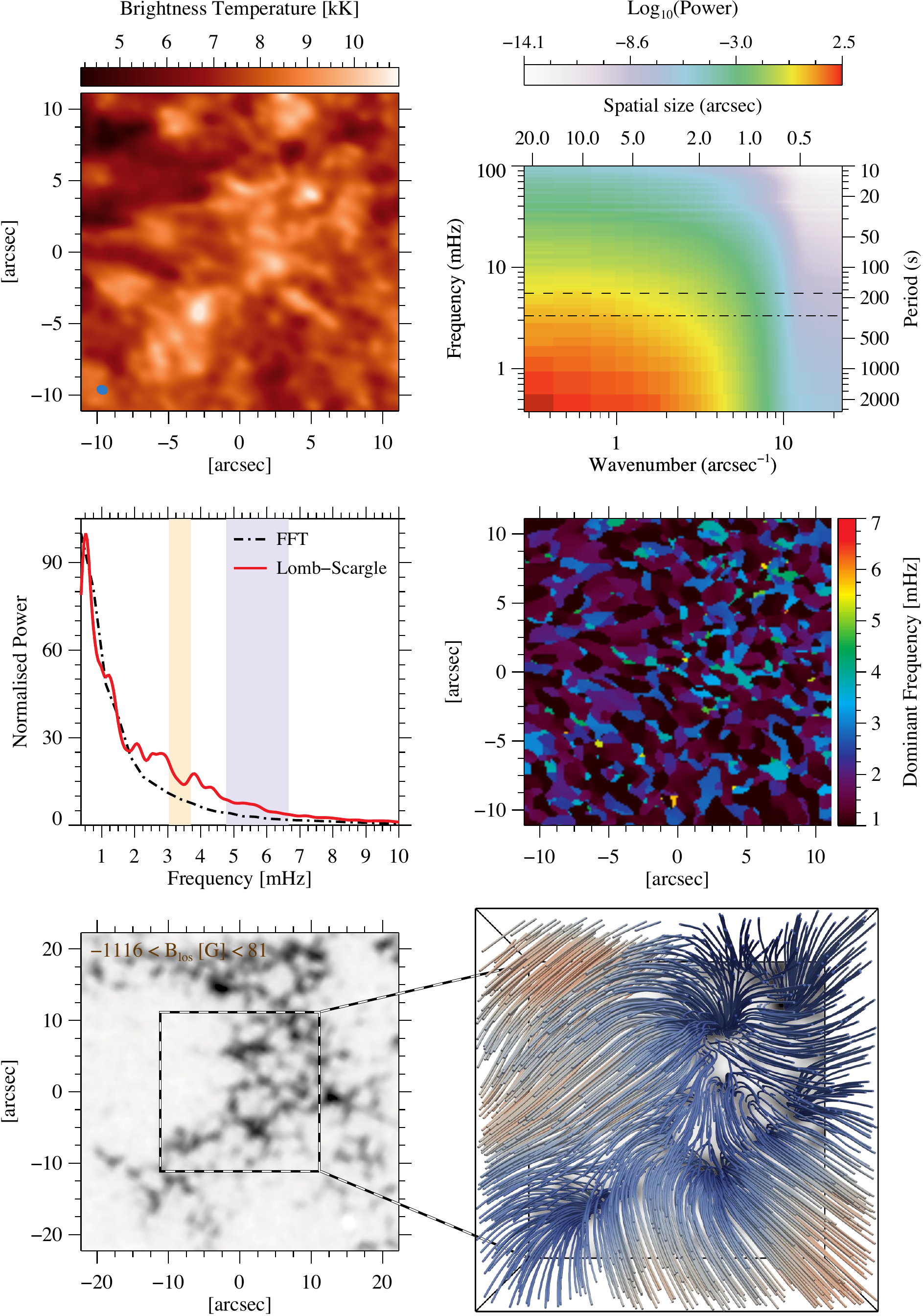}
\caption{
Same as Figure~\ref{fig_D1}, but for ALMA's Band-6 observations from 2017 April 22$^{\text{nd}}$ (i.e., D8 dataset).
}
\label{fig_D8}
\end{figure}

\vspace*{-3pt}
\subsection{D8 dataset: ALMA Band 6, 2017 April 22$^{\text{nd}}$ (Figure~\ref{fig_D8})}

This dataset (also appeared in \citet{2020A&A...634A..56D} and \citet{2020arXiv200512717C}) samples the same plage region as in D2, but here, in Band 6 with a smaller FOV, observed about 80~min earlier than D2/Band-3 (see Table~\ref{table_obslog}). Thus, the field geometries (at around the mid-to-high chromosphere) is also very similar to that of D2, i.e., vertical and near-vertical fields in about half of the FOV, as well as horizontal fields originating in both the plage region within the ALMA's FOV and in those in its immediate vicinity. Similar to D2, although at a different wavelength, the FOV is mostly dominated by low frequencies, smaller than 2~mHz, with only a few insignificant power enhancements within the range of $2-5$~mHz (mostly in structures larger than 15~arcsec). The peak frequencies around 4~mHz are seen at various spatial locations, but mostly around the more vertical fields at chromospheric heights.

\begin{figure}[!htp]
\centering\includegraphics[width=\textwidth]{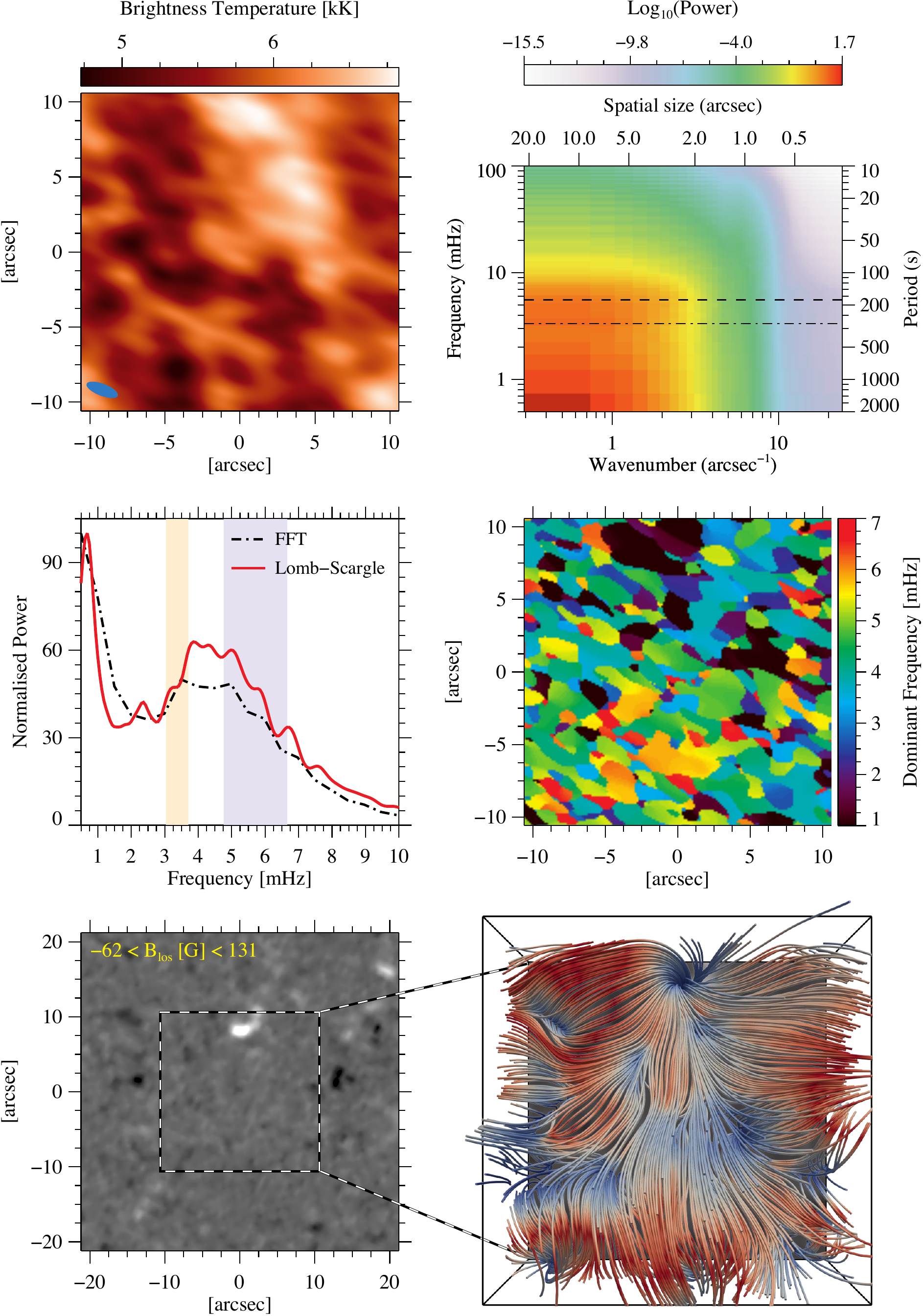}
\caption{
Same as Figure~\ref{fig_D1}, but for ALMA's Band-6 observations from 2018 April 12$^{\text{th}}$ (i.e., D9 dataset).
}
\label{fig_D9}
\end{figure}

\vspace*{-3pt}
\subsection{D9 dataset: ALMA Band 6, 2018 April 12$^{\text{th}}$ (Figure~\ref{fig_D9})}

This observation samples a very quiet (extended) region (i.e., both in ALMA's FOV and its surrounding areas), with only a few, relatively small, field concentrations. This has, consequently, resulted in relatively less dense magnetic canopies whose spatial extents are smaller than those rooted in larger/stronger field concentrations. A very clear power enhancement is found around $3-6$~mHz frequencies, with peak frequencies reaching to higher values of up to 8.3~mHz in several places (the scale of the peak-frequency map is limited to $1-7$~mHz, for the sake of comparison among different datasets).

\begin{figure}[!htp]
\centering\includegraphics[width=\textwidth]{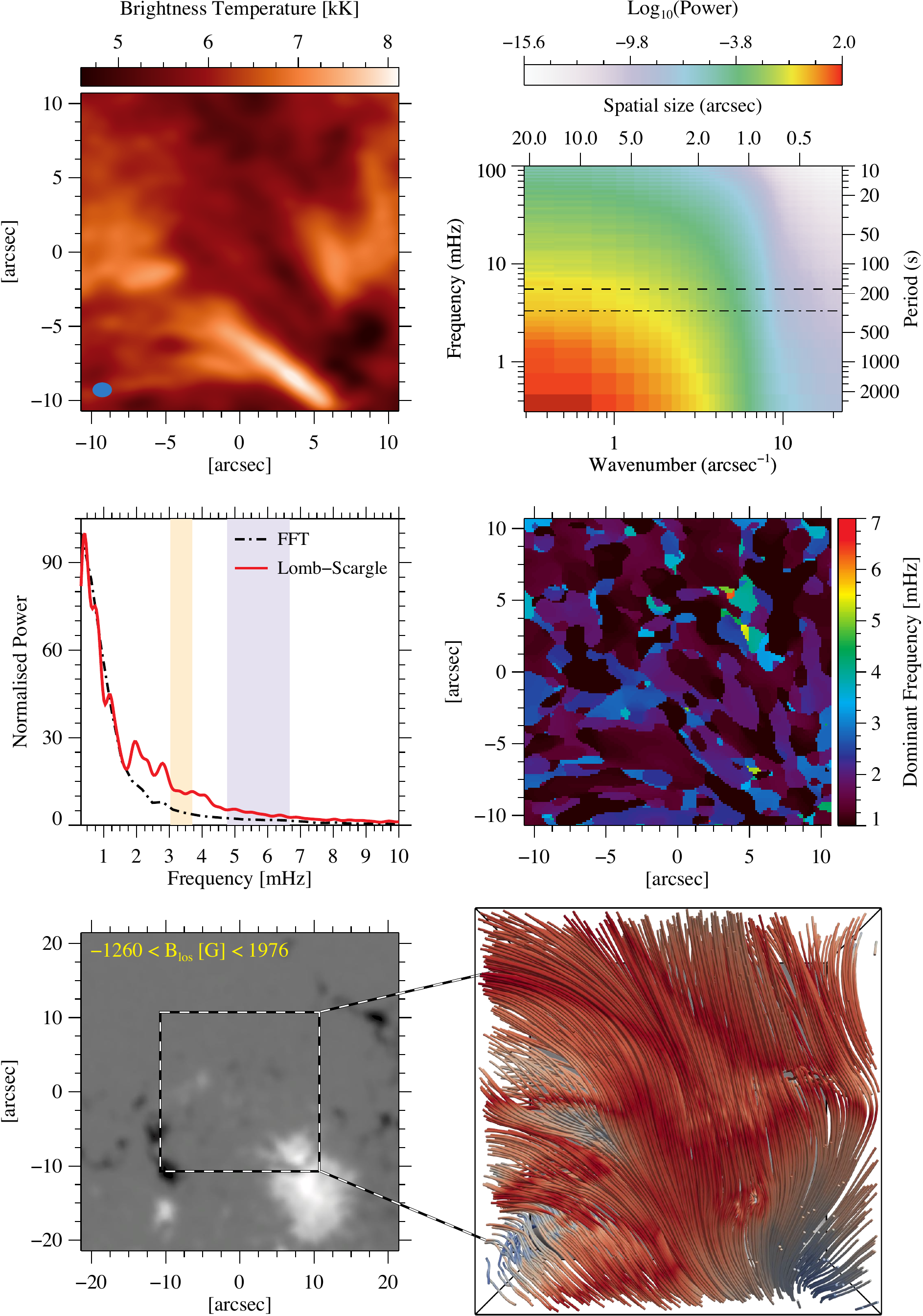}
\caption{
Same as Figure~\ref{fig_D1}, but for ALMA's Band-6 observations from 2018 August 23$^{\text{rd}}$ (i.e., D10 dataset).
}
\label{fig_D10}
\end{figure}

\vspace*{-3pt}
\subsection{D10 dataset: ALMA Band 6, 2018 August 23$^{\text{rd}}$ (Figure~\ref{fig_D10})}

ALMA's FOV includes a few strong field concentrations, i.e., small pores (in the photosphere), mostly around the lower edges, but is more significantly influenced by a large pore just outside the FOV and close to the lower-right corner. Thus, the FOV is very much covered at chromospheric heights by the canopy rooted in the large and small pores. Some field re-configurations are also observed in two small areas, which seem to partly coincide with higher peak frequencies of $3-5$~mHz. Otherwise, the FOV is dominated by low frequencies of ${1-2}$~mHz (mostly at spatial scales larger than 9~arcsec), but also with some power enhancements within $2-3$~mHz.

\section{Discussion and Conclusion}
\label{section_conclusion}

We have studied temperature fluctuations in the solar chromosphere from ALMA (in both Band~3 and Band~6), in a variety of solar regions with different magnetic field strengths and configurations. In addition to the classical FFT technique, we also analysed the non-uniformly sampled image-sequences through the use of Lomb-Scargle approaches. We find, however, that classical FFT and Lomb-Scargle approaches offer similar power spectra at the detected frequencies. We also computed the magnetic field configurations at chromospheric heights, using field extrapolations of the photospheric vector magnetic fields from SDO/HMI. Since the precise formation heights of the ALMA observations are still unknown, one-to-one correlations between the magnetic field geometries and the peak-frequency maps are not straightforward, thus they have only been compared qualitatively (i.e., the field geometries at around $1500-2500$~km, corresponding to the mid-to-high chromosphere, are compared with the ALMA observations in Band~3 and Band~6).

\begin{figure}[!htp]
\centering\includegraphics[width=0.9\textwidth]{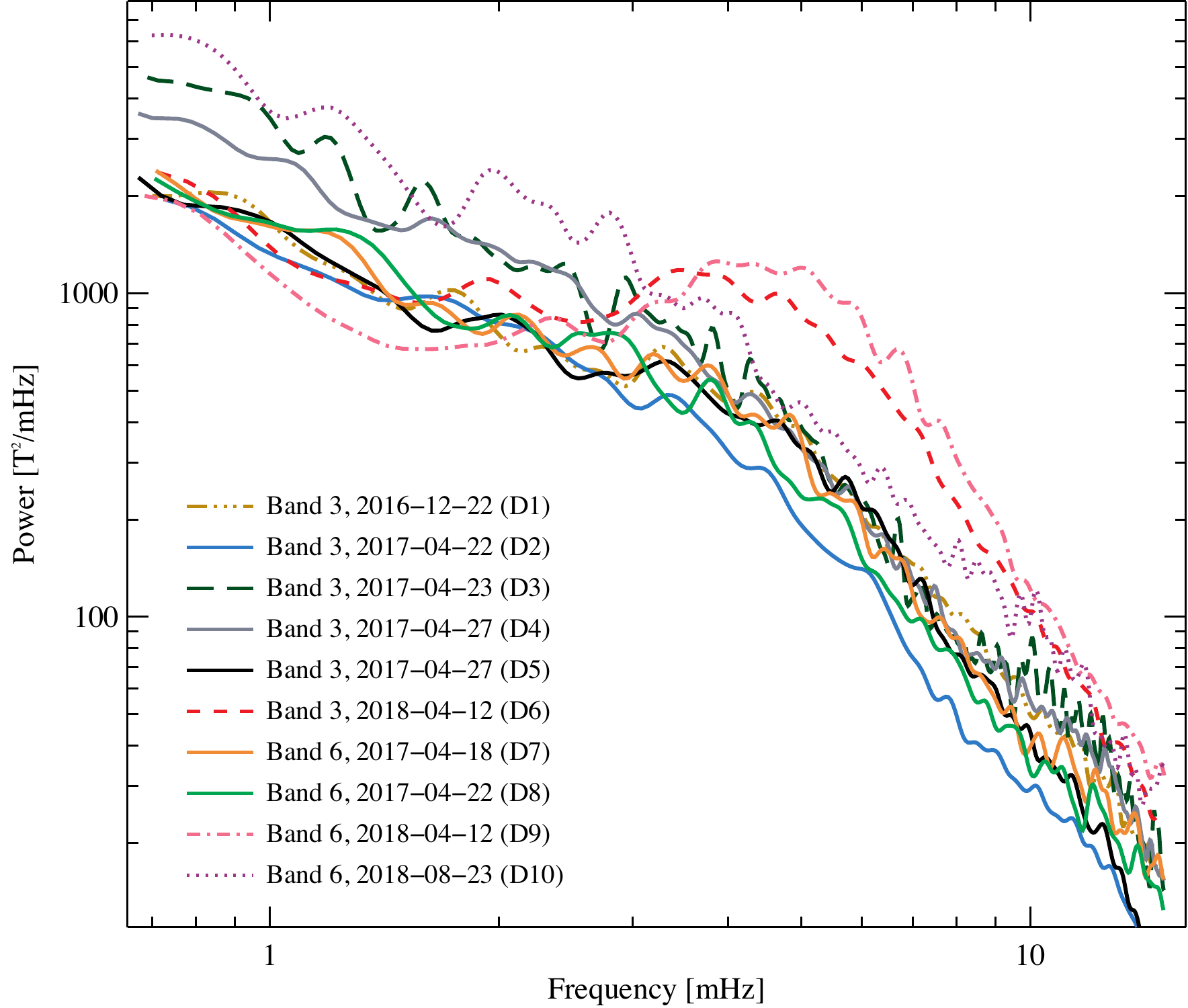}
\caption{Lomb-Scargle power spectra from the ten ALMA datasets D1--D10, as illustrated in Figures~\ref{fig_D1}--\ref{fig_D10}.
}
\label{fig_all_power_spectra}
\end{figure}

The dominant frequencies of the observed oscillations seem to largely depend on the magnetic-field geometries present in the (mid-to-upper) chromosphere, which are derived from the photospheric magnetic configurations within both the observed FOVs and their immediate vicinities. Figure~\ref{fig_all_power_spectra} summarises the mean Lomb-Scargle power spectra (averaged over each entire FOV) for the ten datasets studied here. Only two datasets (i.e., D6 and D9) show clear power enhancements within the frequency range $3-5$~mHz, with a peak at around 4~mHz, whereas the other eight power spectra are peaked at very low frequencies, with occasional small power enhancements at various higher frequencies. The latter high-frequency enhancements (up to 7~mHz) are associated with relatively small numbers of pixels within the FOVs (see the peak-frequency maps in Figures~\ref{fig_D1}--\ref{fig_D10}). We note that although the D6 and D9 datasets were coincidentally observed on the same date, this cannot be the reason for the appearance of power enhancements at around 4~mHz. Instead, this is most likely due to these observations sampling the quietest solar region in our sample. In fact, it was found that these two quiet datasets (i.e., D6 and D9) had smaller surface mean magnetic fluxes, by $1-3$ orders of magnitude (averaged over their entire FOVs; also, similar flux ratios at chromospheric heights), compared to the other eight observational fields of view (see Table~\ref{table_hmi}). Furthermore, the D6 and D9 datasets have the largest oscillatory power compared to the other image sequences. As can be seen in Figure~\ref{fig_all_power_spectra}, this is evidenced by datasets D6 and D9 showing prominent spectral power excesses in the range of $3-5$~mHz, in addition to the largest frequency-integrated power (ranked number 2 and 3 of the 10 datasets), suggesting that these particular datasets have heightened wave energies across the entire frequency spectrum. These findings are consistent with previous coronal studies \cite{2016ApJ...828...89M}, where the largest velocity wave power was found to be correlated with low magnetic flux regions, and in particular, at frequencies larger than 3~mHz.

There have only been a few studies focusing on the dominant frequencies of oscillations in the upper chromosphere. From narrow-band H$\alpha$ line-core image sequences, \citet{2016ApJ...823...45K} found that the oscillatory power of intensity fluctuations (in the high chromosphere), within an evolving active region, was greatly suppressed within the frequency range of $4-8$~mHz (i.e., lack of dominant 3-min oscillations), with some increases found in the $2.7-4.0$~mHz range, and a larger power enhancement at even lower frequencies (i.e., within $1.7-2.7$~mHz). This could be explained by the umbrella effect, where the magnetic canopy covers most of the oscillations in the sub-canopy heights. \citet{2007ApJ...655..624D} reported 3-min oscillations in narrow-band H$\alpha$ line-core time-series of images above two sunspots and in dense plage regions, 5-min fluctuations in adjacent to dense plage regions, and longer periodicities in more inclined-field areas (i.e., the majority of the FOV). Such $3-5$~min oscillations were attributed to dynamic fibrils in and around plage regions. Dynamic fibrils and spicules have also been identified in ALMA Band~6 observations \cite{2020arXiv200512717C}.

In the present work, while the power suppression at frequencies larger than 2~mHz may suggest the presence of such umbrella/canopy effects in the eight datasets with large amounts of magnetic flux, the observations demonstrating significant power enhancements at around $3-5$~mHz in the two most quiescent datasets are in agreement with the propagation of such oscillations throughout the chromosphere. In addition, the magnetic-field inclinations have been shown to play a key role, at different atmospheric heights, in guiding or suppressing the oscillations \cite{2010A&A...510A..41K}. A wide distribution of the inclination angles, from vertical to horizontal fields, were observed at chromospheric heights in the datasets under study. Furthermore, the $p$-mode oscillatory power in the high chromosphere can be suppressed by the presence of strong magnetic fields \cite{2016ApJ...823...45K}, which is the case in each of the eight datasets. Magneto-acoustic waves have also been shown to considerably dissipate their energy by the mid-chromosphere in small, vertical flux concentrations, such as pores, thus, they are much diminished by the high chromosphere \cite{2015ApJ...806..132G,2015A&A...579A..73M,2016ApJ...817...44F,2020arXiv200711594G}. These energy damping waves were, however, observed in relatively small magnetic structures and is likely not responsible for the lack of global oscillatory power at high chromospheric heights, sampled by ALMA observations. Indeed, the global 3-min oscillations have also been observed in the high chromosphere and transition region \cite{2000ApJ...531.1150W}.
It is worth noting that high-frequency MHD waves (on the order of $9-17$~mHz) have also been detected in small-scale structures in ALMA's Band~3 observations (i.e., in the D2 dataset presented here; \cite{Guevara-Gomez2020}). Thus, many of the high-frequency oscillations (up to 9~mHz) we observed in small areas within the FOV of our datasets (see peak-frequency maps in Figures~\ref{fig_D1}--\ref{fig_D10}), could also be due to various types of MHD waves which have vastly been observed in the mid-to-high chromosphere at small scales \cite{2015A&A...577A..17S,2017ApJS..229....7G,2017ApJS..229....9J,2017ApJS..229...10J}.

We cannot completely rule out that the lack of prominent 3-min oscillations is the result of the waves not displaying temperature fluctuations at these frequencies as the waves propagate through the FOV in the upper chromosphere. Under adiabatic conditions, temperature fluctuations are to be expected for magneto-acoustic oscillations (i.e., temperature changes with pressure at constant entropy; \cite{2013SoPh..284..297S}). However, in isothermal atmospheres (where the density and pressure changes take place at constant temperature) the associated wave rarefactions and compressions have a timescale ($\sim1/\omega$) to exchange heat in order to maintain a constant temperature. Isothermal conditions have been found in the solar corona \cite{2011ApJ...727L..32V}, so the possibility of such plasma conditions in the high chromosphere cannot be ruled out. Moreover, it has been suggested that ALMA observations may also have some (small) contributions from coronal heights \cite{2020A&A...635A..71W}. 

Although gravity waves have been suggested to not reach high chromospheric heights (i.e., they are reflected back to the lower atmospheric levels by the magnetic canopy as slow magneto-acoustic waves; \cite{2010MNRAS.402..386N}), the dominant low frequencies of $1-2$~mHz that we find here could also perfectly align with such waves that were previously observed in the lower chromosphere \cite{2008ApJ...681L.125S}. We cannot, however, verify here the nature of the waves represented by the low-frequency ALMA oscillations. 

\begin{figure}[!htp]
\centering\includegraphics[width=0.9\textwidth]{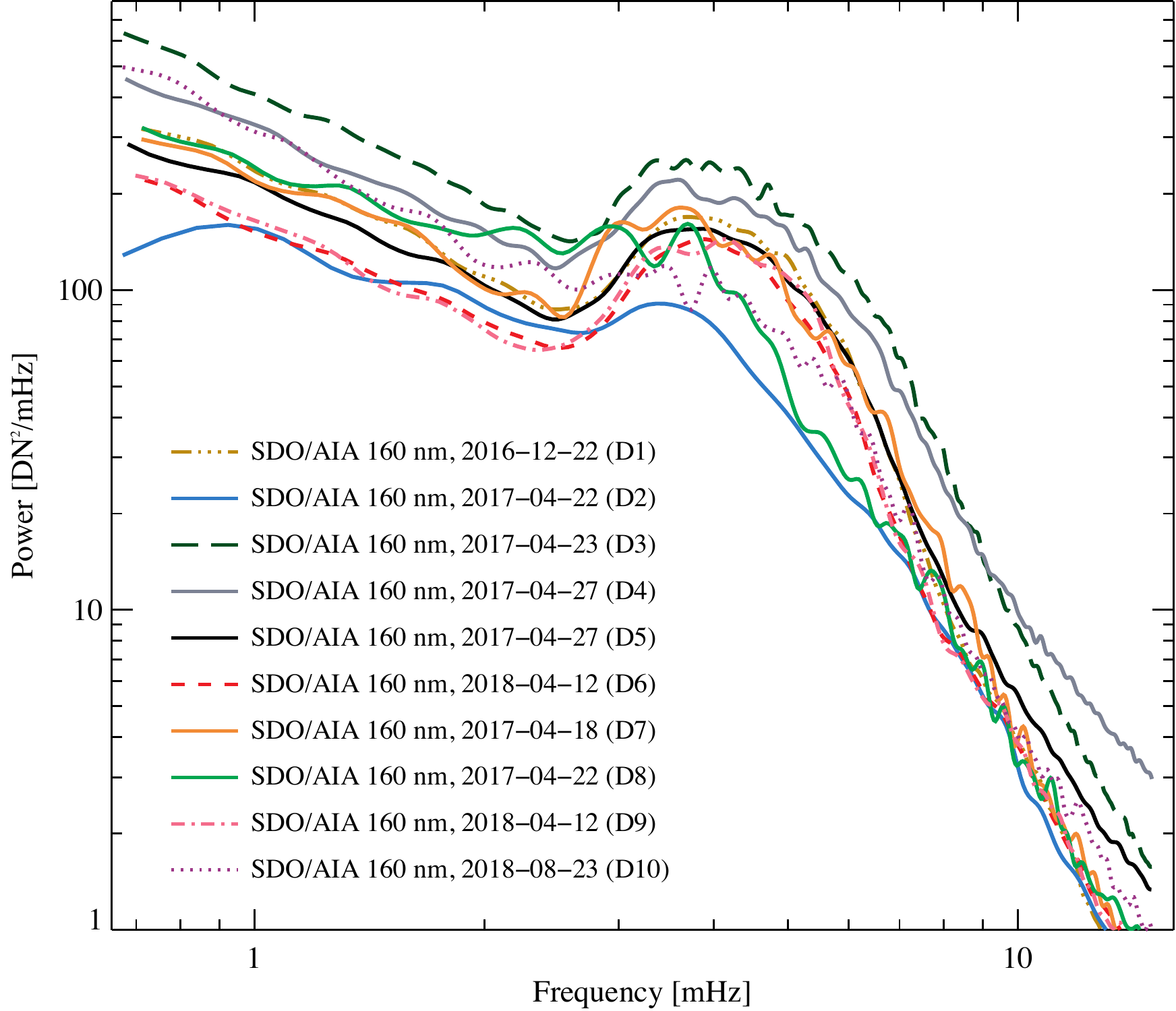}
\caption{Lomb-Scargle power spectra from the SDO/AIA~160~nm time-series of images, corresponding to the ten ALMA datasets D1--D10. To facilitate comparison, the power spectra are plotted with the same line styles and colours as in Figure~\ref{fig_all_power_spectra}.
} 
\label{fig_all_power_spectra_aia1600}
\end{figure}

Considering that all chromospheric diagnostics (except H$\alpha$ intensity) show very pronounced 3-min oscillations dominating the dynamics of the chromosphere, the fact that only a very small fraction of all the pixels in the ten datasets analysed here show peak power near 5.5 mHz may be considered a surprise, in particular since 1-D simulations predicted a very strong, highly nonlinear response in mm temperature brightness to the 3-min oscillations \cite{2004A&A...419..747L,2006A&A...456..713L}. Why do the 1-D hydrodynamic simulations work remarkably well in reproducing other diagnostics such as, e.g., the temporal evolution of the Ca~{\sc ii}~H and K line profile \cite{1992ApJ...397L..59C}, but seem to completely fail in reproducing the mm response to these oscillations? Magnetic fields are certainly an important factor, but they should be a similarly important factor for other diagnostics. 
We should note that the $3-5$~min fluctuations have previously been reported in mm observations (at 85~GHz) from the 10-element Berkeley-Illinois-Maryland Array \cite{2006A&A...456..697W} (with spatial resolution of $\approx10$~arcsec), and from observations with ALMA Band~3 \cite{2020A&A...634A..86P}. In both studies, the authors applied a high-pass filter to the data by subtracting a third-order polynomial fit from the time series. For short time series of just 10~min as that used by \citet{2020A&A...634A..86P}, this filters out oscillations below $\sim$3~mHz, leaving a power peak at around 4~mHz. It has less of an effect for the 30~min long series studied by \citet{2006A&A...456..697W}. In the present study, the signal de-trending was performed by subtracting a simple linear fit, i.e., we did not filter out lower frequencies to visually ``enhance'' power at higher frequencies.
Also, as illustrated in Figure~\ref{fig_all_power_spectra_aia1600}, contemporaneous AIA~160~nm series (co-aligned to the ALMA observations) show a very different distribution in the peak power compared to the ALMA series, with a strong oscillatory signal in the $3-5$~mHz range. Perhaps the answer to this conundrum lies in the great similarity of the ALMA 3~mm brightness temperature maps to H$\alpha$ line width maps, which show a remarkable degree of correlation \cite{2019ApJ...881...99M}. In part, this correlation is driven by significant variations in the height of formation of H$\alpha$ and the 3~mm continuum. Such a strong temporal modulation in the height of formation of the mm continuum could easily destroy the oscillatory signal. 
We also note that no power enhancement is observed in any of AIA~30.4~nm time series (sampling the transition region) corresponding to the ten ALMA datasets.

To summarise, we have analysed the global (average) oscillatory behaviour of temperature fluctuations in ten different ALMA observations (each with different compositions and levels of magnetic flux), recorded in Band~3 and Band~6. We have found a lack of dominant chromospheric oscillations within the frequency range $3-7$~mHz in eight datasets, while there are clear power enhancements at around 4~mHz in the other two (which are the most magnetically quiescent datasets). From field extrapolations of the photospheric magnetic fields, we found that the former datasets were largely influenced by the strong magnetic fields originating within the observed FOVs and/or their immediate vicinities. While we discussed above various possible scenarios to explain these oscillatory behaviours, we conjecture that the lack of 3-min (5.5~mHz) oscillations may be a result of ($a$) the ``umbrella'' effect due to the magnetic canopy, ($b$) power suppression in the presence of strong magnetic fields, ($c$) significant variations in the height of formation, or ($d$) waves not displaying temperature  fluctuations at these frequencies. A larger number of datasets from ALMA and/or observations of the high chromosphere at other wavelengths (i.e., from other instruments) can help provide a better overall picture and clarify the dependency of the oscillation properties on the magnetic geometries.

\enlargethispage{20pt}

\dataccess{The observational data are all publicly available at data archives of ALMA and NASA’s Solar Dynamics Observatory.}

\aucontribute{MS, VMJH, and SJ performed the data reduction and post processing. XZ and TW did the magnetic-field extrapolations. SJ performed the scientific analysis and drafted the manuscript, with assistance from SW, BF, MS, DBJ, RJM, XZ, TW, JCGG, SDTG, BC, KR, and SMW. All authors read and approved the manuscript.}

\competing{The authors declare that they have no competing interests.}

\funding{This work is supported by the SolarALMA project, which has received funding from the European Research Council (ERC) under the European Union’s Horizon 2020 research and innovation programme (grant agreement No. 682462) and by the Research Council of Norway through its Centres of Excellence scheme, project number 262622 (Rosseland Centre for Solar Physics). DBJ and SDTG are supported by an Invest NI and Randox Laboratories Ltd. Research \& Development Grant (059RDEN-1).}

\ack{DBJ and SDTG are grateful to Invest NI and Randox Laboratories Ltd. for the award of a Research \& Development Grant (059RDEN-1). XZ and TW acknowledge support by DFG-grant WI 3211/4-1. BC acknowledges support by NSF grant AGS-1654382 to NJIT. 
We wish to acknowledge scientific discussions with the Waves in the Lower Solar Atmosphere (WaLSA; \href{https://www.WaLSA.team}{www.WaLSA.team}) team, which is supported by the Research Council of Norway (project no. 262622) and the Royal Society (award no. Hooke18b/SCTM).
This paper makes use of several ALMA datasets with the following project numbers: ADS/JAO.ALMA\#2016.1.00423.S, ADS/JAO.ALMA\#2016.1.00050.S, ADS/JAO.ALMA\#2016.1.01129.S, ADS/JAO.ALMA\#2016.1.01532.S, ADS/JAO.ALMA\#2016.1.00202.S, \\ADS/JAO.ALMA\#2017.1.00653.S, and ADS/JAO.ALMA\#2017.1.01672.S.
We are grateful to Bart de Pontieu and Alexander Nindos, PIs of ALMA datasets linked to project numbers 2016.1.00050.S and 2017.1.00653.S, respectively, and for their useful comments. 
ALMA is a partnership of ESO (representing its member states), NSF (USA) and NINS (Japan), together with NRC (Canada), NSC and ASIAA (Taiwan), and KASI (Republic of Korea), in cooperation with the Republic of Chile. 
The Joint ALMA Observatory is operated by ESO, AUI/NRAO and NAOJ. 
We are grateful to the many colleagues who contributed to developing the solar observing modes for ALMA and for support from the ALMA Regional Centres. 
The AIA and HMI data are courtesy of the NASA/SDO, as well as AIA and HMI science teams.
}


\bibliographystyle{rstasj}
\bibliography{references}

\end{document}